\documentclass[aps,twocolumn,prb,showpacks,preprintnumber]{revtex4-1}
\usepackage[dvips]{graphicx}
\usepackage{dcolumn}
\usepackage{amsmath,amssymb}
\usepackage{txfonts}
\usepackage{bm}

\begin{document}
\draft
\title{Reduced Hamiltonian for Electronic States of Dilute Nitride Semiconductors}

\author{Masato Morifuji$^1$ and Fumitaro Ishikawa$^2$}

\affiliation{$^{1}$Graduate School of Engineering, Osaka University, 2-1 Yamada-oka, 
Suita, Osaka 565-0871, Japan\\
$^{2}$Graduate School of Science and Engineering, Ehime University, 3 Bunkyo-cho, 
Matsuyama, Ehime 790-8577, Japan}

\date{\today}

\begin{abstract}
We present a novel model to describe conduction band of GaN$_x$As$_{1-x}$ (GaNAs).
As well known, GaNAs shows exotic behavior such as large band gap bowing.
Although there are various models to describe the conduction band of GaNAs, 
origin of the band gap bowing is still under debate.
On the basis of perturbation theory, we show that the behavior of conduction band  
is mainly arising from intervalley mixing between $\Gamma$ and L or X.
By using renormalization technique and group theoretical treatment, 
we derive a reduced Hamiltonian which describes well the band gap shrinkage of GaNAs.
\end{abstract}

\pacs{}
\maketitle

\section{Introduction}

III-V compound semiconductors containing nitrogen have been extensively studied  for their 
properties different from those of conventional 
semiconductors.\cite{Kondow,Weyers,Bi,Skierbiszewski,Shan1,Tan,Uesugi}
In particular, behavior of conduction band edge of GaN$_x$As$_{1-x}$ (GaNAs)
with small $x$ attracts wide attention.\cite{Noguchi,Sumiya,Fukushima}
As well known, the band gap of GaNAs decreases with nitrogen concentration.
This is contrary to the conventional Begard's law, that is, band gap of a mixed compound
 is well described as  a linear interpolation of band gaps of constituent materials.

There have been various  models to explain such  behavior of  GaNAs.\cite{Zhao,Shan2,Wu}
However, origin of the band gap bowing is still under debate.
Band anticrossing model \cite{Shan2,Wu} is widely used for phenomenological 
explanation of experimental results,
however, its physical foundation is ambiguous.
Band theories, which is a powerful tool to investigate electronic states, 
also have been applied to GaNAs.
The tight-binding model, \cite{Lindsay,OReilly,Fan} 
empirical pseudopotentials, \cite{Bellaiche,Bellaiche2,Mader,Kent}
and the first principle calculations \cite{Timoshevskii,Tan2}  have been carried out to reproduce  
experimental results.
Reliable results were obtained from these calculations, 
however, physical insight can be missed in handling the large matrix containing all the effects.
In addition, when nitrogen concentration is very low, band calculations require 
much computational resources. 
As a result, calculations become difficult to carry out.


In  this study, in order to investigate the conduction band of GaN$_x$As$_{1-x}$ with small $x$,
 we present a novel model  derived from  perturbation calculations 
using  wavefunctions of bulk GaAs as bases.
Investigation on behavior of conduction band 
has revealed  that  intervalley mixing 
induced by lattice distortion around nitrogen plays an important role 
for the band gap reduction of GaNAs.
Utilizing this result, along with renormalization technique and symmetry considerations, 
we derive a simple equation to evaluate energy of the conduction band of GaNAs.

\begin{figure}[h]
\includegraphics[width=7.0cm]{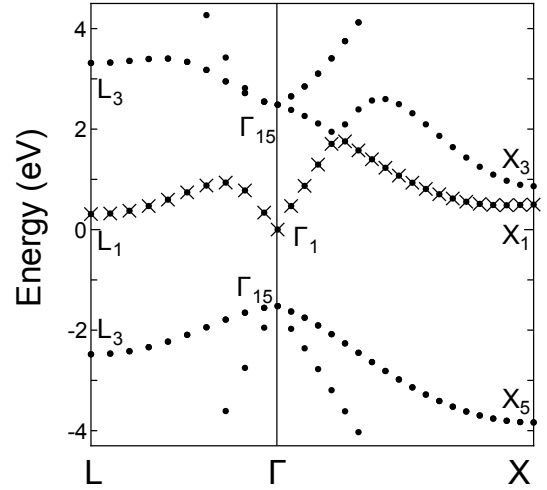}
\caption{Dispersion curves of bulk GaAs calculated by using plane wave bases and empirical pseudopotentials.
The lowest conduction band, which we consider in this study, is denoted by crosses.
Origin of energy axis is set to the conduction band edge.}
\label{fig:1}
\end{figure}

\section{Theory}

\subsection{Overview of band calculation procedure}

First, we briefly review the procedure to calculate electronic states of bulk GaAs 
within the empirical pseudopotential method \cite{Bellaiche}
to be used as the basis in the following calculations.

Hamiltonian of bulk GaAs is given by a summation of kinetic energy 
and atomic potential energy $V({\bm r})$ as
\begin{align}
{\cal H}_0=-\frac{\hbar^2\nabla^2}{2m}+V({\bm r}),
\label{eq:1}
\end{align}
with
\begin{align}
V({\bm r})=\sum_{i}\left[ V_{\rm Ga}({\bm r}-{\bm \tau}-{\bm R}_i)
         + V_{\rm As}({\bm r}-{\bm R}_i)\right],
\label{eq:2}
\end{align}
where $V_{\rm Ga}\, (V_{\rm As})$ is atomic pseudopotential of Ga (As) located in a unit cell 
specified by a lattice vector of the zinc blende structure ${\bm R}_i$.
$\bm \tau=(a/4,a/4,a/4)$ with $a$ the lattice constant is a vector 
which specifies position of Ga within a unit cell.
Based on the empirical point of view, 
we regard that Coulomb interaction between electrons, exchange and correlation interactions, etc. 
are effectively included in the atomic pseudopotentials.
We neglected the spin-orbit interaction.
First, we calculate a Hamiltonian matrix 
\begin{align}
{\cal H}_{0\, \bm k+\bm G_i, \bm k+ \bm G_j}\equiv \langle{\bm k+\bm G_i}|{\cal H}_0|{\bm k+ \bm G_j} \rangle, 
\label{eq:3}
\end{align}
 using plane wave basis functions 
\begin{align}
\langle \bm r\,|\,\bm k + \bm G_i\rangle =\frac{1}{\sqrt{\Omega}}\,e^{i({\bm k+\bm G_i)}\cdot{\bm r}},
\label{eq:4}
\end{align}
with
$\bm k$ a wavevector, $\bm G_i$ a reciprocal lattice vector, 
and  $\Omega$ the system volume, respectively.
By diagonalizing the Hamiltonian matrix, we can  calculate a band energy
\begin{align}
\varepsilon^{0}_{n,\bm k}=
\langle\psi^{0}_{n, \bm k}|{\cal H}_0|\psi^{0}_{n, \bm k'}\rangle\,\delta_{\bm k,\bm k'},
\end{align}
and a wavefunction 
\begin{align}
\psi^{0}_{n, \bm k}({\bm r})=\sum_i c_{n, \bm k+\bm G_i}|\bm k + \bm G_i\rangle,
\label{eq:5}
\end{align}
where $c_{n,\bm k+ \bm G_i}$ is an eigenvector with an index $n$ specifying band.
The superscript \lq\lq 0\rq\rq indicates non-perturbed quantities.
In Figure~\ref{fig:1}, we show dispersion curves of bulk GaAs 
evaluated using empirical pseudopotential,\cite{Bellaiche}
where zero of the energy axis is set to the conduction band edge.

Using the bulk wavefunctions, we carry out perturbation calculations to 
evaluate energies of GaNAs.
In what follows, we consider only the lowest conduction band plotted by crosses 
because we are interested in behavior of the conduction band edge labeled by $\Gamma_1$ in Fig.~\ref{fig:1}.
From now on, we thus omit the index $n$ which specifies band.

\subsection{Perturbation matrix}

Let us  consider an $N \times N \times N$ supercell in which  one of As atoms therein 
 is replaced by a nitrogen atom.
Although it is possible to apply the present theory for a system containing many nitrogen atoms,
 in this paper, we treat only the case of a single nitrogen atom.
This supercell contains $4N^3$ primitive cells of the zinc blende structure.

Introduction of an N atom gives rise to change in crystalline potential.
We  take three factors into account:
(i) change of the atomic potential from As to N, (ii) displacement of Ga atoms neighboring to the N atom, 
and (iii) displacement of As atoms at the second neighboring positions to the N atom.
Then, the perturbation Hamiltonian is written as
\begin{align}
{\cal H}'({\bm r})
       &=\left[V_{\rm N}({\bm r}-{\bm R}_I)-V_{\rm As}({\bm r}-{\bm R}_I)\right] \notag \\
       &+\sum_j\left[V_{\rm Ga}({\bm r}+{\bm \tau}-{\bm R}_j-{\bm \xi}_j)
               -V_{\rm Ga}({\bm r}+{\bm \tau}-{\bm R}_j) \right]\notag \\
       &+\sum_{j'}\left[V_{\rm As}({\bm r}-{\bm R}_{j'}-{\bm \eta}_{j'})
                -V_{\rm As}({\bm r}-{\bm R}_{j'})\right],
\label{eq:6}
\end{align}
In the right hand side of eq.~(\ref{eq:6}), the first term denotes potential change 
from that of As to N located at the position ${\bm R}_I$.
The second and the third terms are arising from 
 displacement of atoms neighboring to the  nitrogen
where $\bm\xi_j$ and $\bm\eta_j$ are the displacement of the first neighboring 
Ga and the second neighboring As, respectively.
The indices $j$ and $j'$ run through so that ${\bm R}_j-{\bm \tau}+{\bm \xi}_j$ and 
${\bm R}_{j'}+{\bm \eta}_{j'}$ indicate 
the positions of the first neighboring four Ga atoms 
and the second neighboring twelve As atoms, respectively.
We set ${\bm \xi}_j$ so that the Ga atoms approach to the N atom by 0.38 \AA.
Similarly, ${\bm \eta}_{j'}$ was determined so that the second neighboring As atoms approach to 
the N by 0.1 \AA.
These values of atom displacements were determined from total energies evaluated 
by the first principle calculations using CASTEP package.

\begin{figure}
\includegraphics[width=8cm]{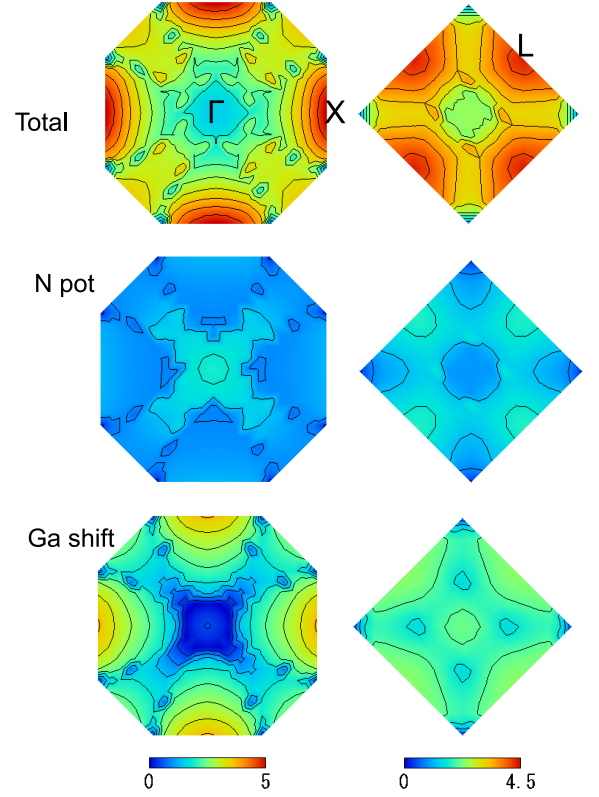}
\caption{(Color Online) Matrix element $|{\cal H'}_{\bm k',\bm k}'|$ with $k'=(0,0,0)$ is plotted 
as a function of $\bm k$ 
on the $k_z=0$ 
plane (left column) and $k_z=\pi/a$ plane (right column) of the first Brillouin zone.
From the top to bottom, total value of the matrix elements, contribution from the factor (i), and 
contribution from the factor (ii) are plotted in the unit of eV.}
\label{fig:2}
\end{figure}

\begin{figure}
\includegraphics[width=8cm]{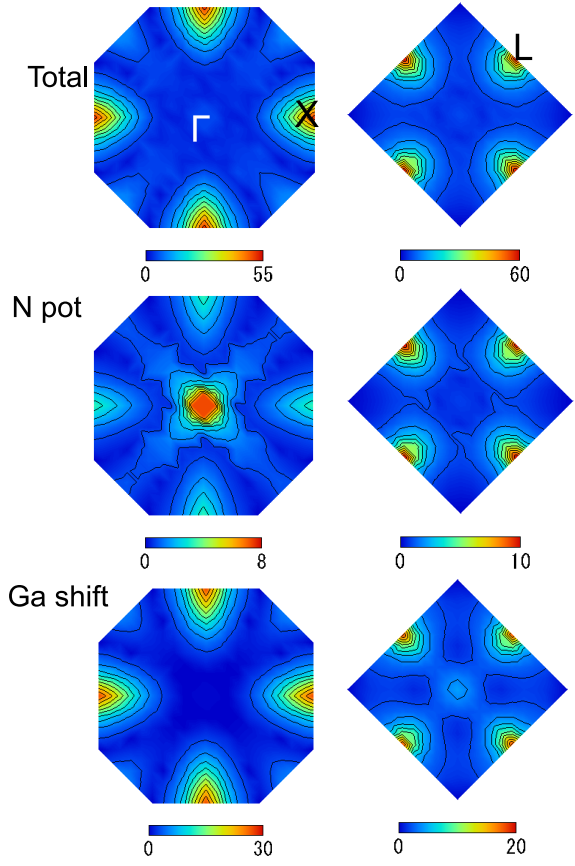}
\caption{(Color Online) $|{\cal H'}_{\bm k',\bm k}|^2/|\varepsilon^{0}_{\bm k'} -\varepsilon^{0}_{\bm k}|$ 
with $k'=(0,0,0)$ is plotted on the $k_z=0$ 
plane (left column) and $k_z=\pi/a$ plane (right colum) of the first Brillouin zone.
From the top to bottom, total value, contribution from the factor (i), and 
contribution from the factor (ii) are plotted.}
\label{fig:3}
\end{figure}

We calculated matrix elements of the perturbation Hamiltonian 
between the Bloch states of bulk GaAs taking the three factors (i), (ii), and (iii) 
mentioned above into account.
In Fig.~\ref{fig:2}, we plot absolute value of the matrix elements 
\begin{align}
{\cal H'}_{\!\!\bm k',\bm k}\equiv \left(\frac{\Omega}{\Omega_{uc}}\right)\,
           \langle \psi^{0}_{\bm k'}|{\cal H'}|\psi^{0}_{\bm k}\rangle,
\label{fig:4a}
\end{align}
for $\bm k' = \Gamma$ as a function of $\bm k$.
$\Omega_{uc}=a^3/4$ is the zinc blende unit cell volume. 
On the left column, $|{\cal H'}_{\Gamma,\bm k}|$s are
 plotted on the $k_z =0$ plane which contains the $\Gamma$ and X-points.
On the right column, $|{\cal H'}_{\Gamma,\bm k}|$s on the $k_z=\pi/a$ plane
 (the L-point is included) are shown.
Note that different scales are used for figures in the left and right columns
and that the values are in unit of eV.
From top to bottom, total value, 
contribution from the factor (i), and contribution from the factor (ii) are plotted, respectively.
We do not show contribution from the factor (iii)  the position shift of second neighboring As,
since this is much smaller than others.
We note that the matrix elements are basically negative values, 
although we plot absolute values since they are complex quantities. 
We see that 
$|{\cal H'}_{\Gamma,\bm k}|$ takes a large value 
when $\bm k$ is X  and L.
We also see that the effect of displacement of neighboring Ga atoms is larger than that of 
the N potential.

Fig.~\ref{fig:3} shows the quantity, 
$|{\cal H'}_{\Gamma,\bm k}|^{2}/|\varepsilon^{0}_\Gamma -\varepsilon^{0}_{\bm k}|$
plotted as a function of $\bm k$.
Similar to  Fig.~\ref{fig:2}, total value, contribution from nitrogen potential, 
and displacement of Ga atoms   are plotted from top to bottom.
It is seen that 
$|{\cal H'}_{\Gamma,\bm k}|^{2}/|\varepsilon^{0}_\Gamma -\varepsilon^{0}_{\bm k}|$
 is largest when $\bm k$ is the L-state, although the matrix element for the X-state 
is larger than that for the L-state.
This is  because energy difference between 
the L and $\Gamma$, $|\varepsilon^{0}_\Gamma -\varepsilon^{0}_L|$, is smaller than $|\varepsilon^{0}_\Gamma -\varepsilon^{0}_X|$.
We also see that the N potential gives rise to mixing between the $\Gamma$-state and states in vicinity 
of $\Gamma$,
whereas displacement of Ga atoms gives rise to the intervalley mixing.
These results indicates that mixing between the $\Gamma$ and L-states and/or 
between the $\Gamma$ and X-states  is relevant to the band gap reduction.

\begin{figure}
\includegraphics[width=8cm]{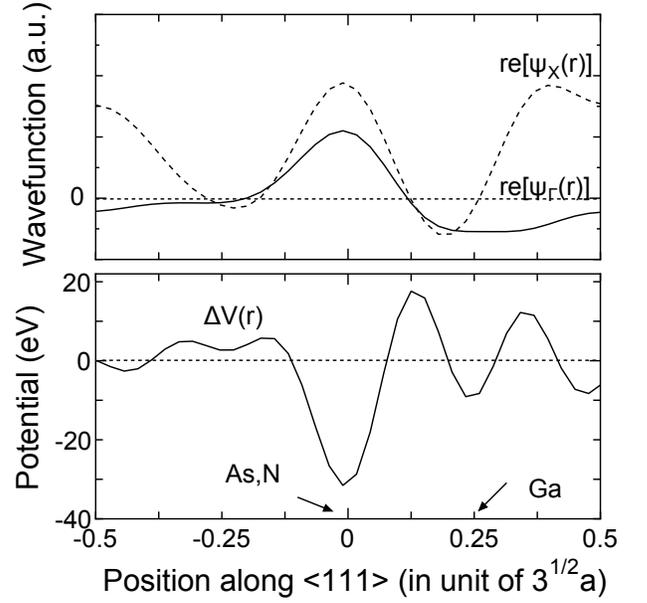}
\caption{Upper panel: wavefunctions of the $\Gamma$-state and X-state are plotted along 
the $\langle 111\rangle$ direction by solid and dashed curves, respectively.
Lower panel: Perturbation potential along the  $\langle 111\rangle$ direction 
is plotted by solid curve.
The arrows show positions of As (N) and Ga.
The horizontal dotted line shows zero.}
\label{fig:4}
\end{figure}

\subsection{Character of wavefunctions and matrix elements}

We can discern 
the $\bm k$-dependence of the perturbation matrix elements
 from wavefunctions of bulk GaAs.
In the upper panel of Fig.~\ref{fig:4}, the solid and dashed curves show 
$\psi^{0}_\Gamma(\bm r)$ and $\psi^{0}_X(\bm r)$ 
plotted along the $\langle 111 \rangle$ direction.
The As (or N) atom locates at the position $0.0$, and  
 a Ga atom without displacement locates at $+0.25$ as indicated by arrows.
It is seen that the $\Gamma$-state wavefunction consists of anti-bonding coupling 
between an $s$-like orbital of As and an $s$-like orbital of Ga.
We also observe that the wavefunction around Ga is largely extended.
The X-state consists of anti-bonding coupling between $s$-like orbital of As and $p$-like orbital of Ga
 which has a node at the Ga position.
The L-state has character similar to the X-state though it is not shown in the figure.

In the lower panel, we plot perturbation potential along the $\langle 111 \rangle$ direction.
We observe that the N atom gives rise to negative potential with $s$-like symmetry.
On the other hand, perturbation potential around the Ga position 
is anti-symmetric around the Ga, that is, $p$-like symmetry.

These curves of crystalline potential and wavefunctions enable us to make 
qualitative interpretation on the 
matrix elements shown in the previous section.
First, we consider the diagonal element ${\cal H'}_{\Gamma,\Gamma}$
This quantity is arising mainly from N potential
 because $\psi^{0}_{\Gamma}$ has a large amplitude at the N position.
On the other hand, shift of Ga contribute little to ${\cal H'}_{\Gamma,\Gamma}$.
This is because $\psi^{0}_{\Gamma}$  has $s$-like character around Ga.
As we have noted, potential change due to Ga displacement is of $p$-character. 
Integration $|\psi^{0}_{\Gamma}|^2 \times \varDelta V$around the Ga atom 
will make the matrix element small.
The coupling between $\Gamma$ and L is also determined in the similar mechanism.

For the coupling between the $\Gamma$- and X-states ${\cal H'}_{\Gamma,X}$,
shift of Ga atoms has an important role.
From Fig.~\ref{fig:4}, we see that $\psi^{0}_{\Gamma}$ has $s$-like symmetry around the Ga atom, 
whereas both $\psi^{0}_{X}$ and $\cal H'$have $p$-like symmetry. 
Therefore, we anticipate that multiplication of these three quantities 
becomes even function around Ga, which 
enlarges the matrix element ${\cal H'}_{\Gamma,X}$.

We noted that contribution from shift of the second neighboring As atoms is small. 
This is also understood from symmetry.
Although displacement of As atoms is about 1/3 of that of Ga atoms, 
contribution might be large because of larger number (12) of neighboring As atoms.
As we have noted, atom's position change gives rise to perturbation potential with $p$-like 
symmetry. 
As seen from Fig.~\ref{fig:4}, 
both the $\Gamma$-state and X-state have $s$-like charge distribution around As atoms.
From a simple consideration on symmetry, 
we see that $\langle\psi^{0}_{\bm k}|{\cal H'}|\psi^{0}_{\bm k'}\rangle$
 with $\bm k$ and $\bm k'$ $\Gamma$ or X has a small value.

These results indicates that mixing between $\Gamma$ and X or between $\Gamma$ and L 
induced by lattice distortion around N gives rise to band gap reduction of GaNAs.

\begin{figure}[h]
\includegraphics[width=7.5cm]{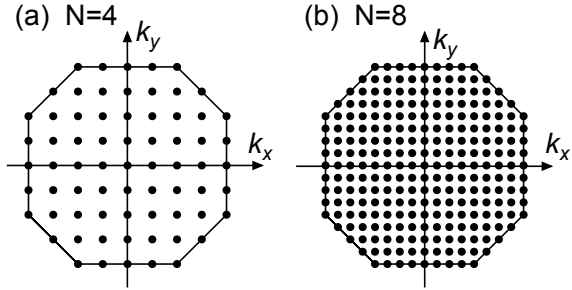}
\caption{Reciprocal lattice vectors used in perturbation calculations  
for (a) $4\times4\times4$ and (b) $8\times 8 \times 8 $ supercells.
Note that equivalent points are excluded in the following calculations 
though they are plotted in the figures.}
\label{fig:5}
\end{figure}

\subsection{Band gap shrinkage} 

The matrix elements of the  perturbation Hamiltonian for a single N atom 
in an $N\times N\times N$ supercell are written as
\begin{align}
{\cal H'}_{\bm k,\bm k'}^{(N)}=\frac{1}{4N^3}\langle \psi^{0}_{\bm k}|{\cal H'}|\psi^{0}_{\bm k'}\rangle,
\label{eq:7}
\end{align}
where $1/{4N^3}$ is a factor to be normalized over the supercell.
The states $\bm k$ and $\bm k'$ at which ${\cal H'}_{\bm k,\bm k'}^{(N)}$ is evaluated 
are obtained as follows:
Since the perturbation potential has translational symmetry with a period $Na$ 
in all the $x$-, $y$-, and $z$-directions, 
$\langle \psi^{0}_{\bm k}|{\cal H'}|\psi^{0}_{\bm k'}\rangle$ 
must be unchanged when  ${\cal H'}({\bm r})$ is replaced by 
${\cal H'}({\bm r+ \bm R})$ with a lattice vector of the supercell $\bm R$.
From this, the wavevectors $\bm k$ and $\bm k'$ in eq.~(\ref{eq:7}) must satisfy a relation
\begin{align}
{\bm k}-{\bm k'}=\frac{2\pi}{Na}(n_x,n_y,n_z),
\label{eq:8}
\end{align}
with $n_x, n_y$, and $n_z$  integers.
In Figs.~\ref{fig:5} (a) and (b), the dots indicate possible $\bm k -\bm k'$ plotted 
on the first Brillouin zone of the zinc blende structure for $N=4$ and  $N=8$, respectively.
Note that some points on the border are equivalent.
For example, $2\pi/a(1,0,0)$ and $2\pi/a(-1,0,0)$ are identical and thus one of them must be excluded,
though both are plotted in the figure.
Excluding such equivalent points, we have $4N^3$ $\bm k$-points in the first Brillouin zone
to be mixed due to the perturbation potential;  
 there are 256 points for $N=4$ and 2048 points  for $N=8$ necessary for calculations.
We also note that these $\bm k$-points are the points that are folded onto the $\Gamma$-point in the 
Brillouin zone of the $N \times N \times N$ supercell.
We can calculate conduction band energy from 
${\cal H'}_{\bm k,\bm k'}^{(N)}$ given by eq.~(\ref{eq:7}) 
with bulk GaAs states shown in Fig.~\ref{fig:5}.

\begin{figure}
\includegraphics[width=8cm]{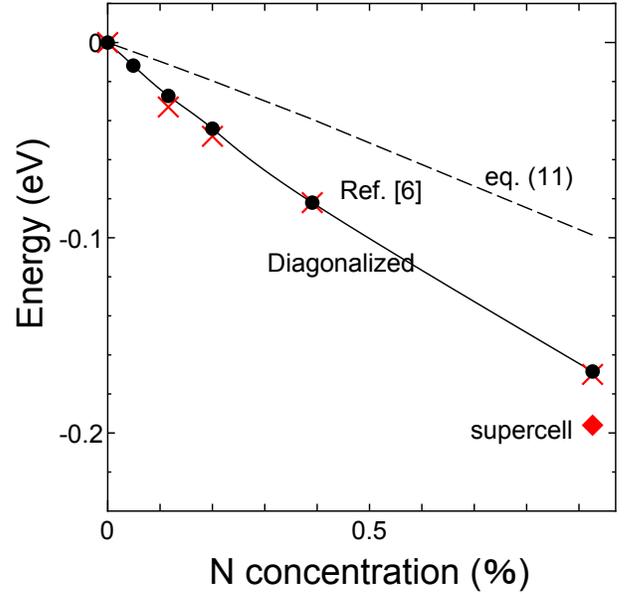}
\caption{(Color Online)  Energies of conduction band edge are shown.
Dashed curves show results of second order perturbation, evaluated by eq.~(\ref{eq:11}).
Filled circles show results calculated by diagonalizing full Hamiltonian matrix given by 
eq.~(\ref{eq:12}).
For comparison, theoretical data from Ref.~[6] and results of supercell calculation 
are also plotted by cross and square, respectively.}
\label{fig:6}
\end{figure}

We may evaluate energy of the conduction band edge 
$\varepsilon_{\Gamma}$ using the matrix elements
by perturbation expansion.
Up to the second order term, the energy change is given by 
\begin{align}
\varepsilon_{\Gamma}=\frac{1}{4N^3}\langle\psi^{0}_\Gamma |{\cal H'}|\psi^{0}_\Gamma \rangle 
                 +\frac{1}{(4N^3)^2}\sum_{\bm k}^\prime 
    \frac{|\langle\psi^{0}_\Gamma |{\cal H'}|\psi^{0}_{\bm k} \rangle|^2}
{\varepsilon^{0}_\Gamma-\varepsilon^{0}_{\bm k}},
\label{eq:11}
\end{align}
where $x=1/4N^3$ is nitrogen concentration.
As we show in Fig.~\ref{fig:6} by dashed curve, 
the second order perturbation is insufficient to explain experimental results, 
indicating that higher order perturbation energies are necessary since 
the potential change due to nitrogen is far from moderate.

Then, in order to evaluate energy of conduction band edge, we diagonalized a matrix
\begin{align}
{\cal H}_{\bm k,\bm k'}^{(N)} \equiv 
\langle\psi^{0}_{\bm k}|{\cal H}_0|\psi^{0}_{\bm k'} \rangle + {\cal H'}_{\bm k,\bm k'}^{(N)}.
\label{eq:12}
\end{align}
Results are shown in Fig.~\ref{fig:6} by filled circles.
For comparison, we plot theoretical data from Ref.~[6] by crosses.
We also plot a result of supercell calculation with cutoff energy 3.0 Ryd.
We see that the present perturbation calculations yield reasonable results.

\section{Reduced Hamiltonian}

As we have shown in the previous section, 
proper energies were evaluated by diagonalizing a Hamiltonian matrix of $4N^3 \times 4N^3$ size.
Mixing between the $\Gamma$-state and other $\bm k$-states 
due to nitrogen doping gives rise to the reduction of the conduction band edge. 
Among the states, the mixing between $\Gamma$ and L is the largest.
This fact leads us to an idea that we may have an effective Hamiltonian with 
only the $\Gamma$ and L as bases.
For this purpose, we applied L\"odin's theory \cite{Loedin}
described in detail in appendix A. 
In the present case where bases are the $\Gamma$ and four L-states,  
we can reduce the size of the  perturbation Hamiltonian down to  $5 \times 5$ as 
shown in eqs.~(\ref{eq:A8}) and (\ref{eq:A9}).

Further reduction of Hamiltonian is possible by applying group theoretical consideration.
Since the four L-states are degenerated, any linear combinations among them 
satisfy the Schr\"odinger equation.
This fact allows us to make a suitable combination that mixes (or does not mix) 
with the $\Gamma$-state.
Coefficients for such a linear combination among the L-states are 
obtained considering symmetry of the L-state.
Following standard procedure to obtain normal modes of an irreducible representation
of the point group T$_d$, we symmetrize the linear combinations among 
the four L-states, $L_1 \sim L_4$.\cite{Burns,Jones}
In this way, we have a singlet state 
\begin{align}
\psi_{\tilde{L}}= \frac{1}{2}\,\psi_{L_1}+\frac{1}{2}\,\psi_{L_2}
+\frac{1}{2}\,\psi_{L_3}+\frac{1}{2}\,\psi_{L_4},
\label{eq:13}
\end{align}
and triplet states
\begin{align*}
&\psi_{L'}= \frac{1}{\sqrt{2}}\,\psi_{L_2}-\frac{1}{\sqrt{2}}\,\psi_{L_3},  \notag\\
&\psi_{L''}= -\frac{1}{\sqrt{2}}\,\psi_{L_1}+\frac{1}{\sqrt{2}}\,\psi_{L_4,} \notag \\
&\psi_{L'''}= \frac{1}{2}\,\psi_{L_1}-\frac{1}{2}\,\psi_{L_2}
            -\frac{1}{2}\,\psi_{L_3}+\frac{1}{2}\,\psi_{L_4}.
\end{align*}
However, we have to note that these coefficients depend on 
situations such as nitrogen position, 
choice of origin, and trivial phase of wavefunctions  etc.
It is necessary to consider situations carefully in evaluating the coefficients.

Since the singlet state given by eq.~\ref{eq:13} connects with the $\Gamma$-state and the triplet states do not,
we can transform the 5$\times$5 reduced matrix into a form
\begin{align}
U^\dagger\, \tilde{\cal H}_{\bm k,\bm k'}^{(N)}\, U
=
\left[
\begin{array}{cc|ccc}
\tilde{\varepsilon}^{0}_{\Gamma} & \tilde{V}_{\Gamma L} & 0 & 0 & 0  \\[1mm]
\tilde{V}_{\Gamma L}^* &\tilde{\varepsilon}^{0}_{L} -3|\tilde{V}_{L L}|& 0 & 0 & 0  \\[0.3mm]\hline \\[-3mm]
 0 & 0     & \tilde{\varepsilon}^{0}_L+|\tilde{V}_{L L}| &     0          &     0        \\[1mm]
 0 & 0     & 0 & \!\!\!\!\!\tilde{\varepsilon}^{0}_L +|\tilde{V}_{L L}|&     0         \\[1mm]
 0 & 0     & 0 & 0   & \!\!\!\!\!\tilde{\varepsilon}^{0}_L+|\tilde{V}_{L L}|
\end{array}
\right],
\end{align}
with a matrix $U$ in the form 

\begin{align}
U=
\left[
\begin{array}{c|cccc}
1  & 0 &  0 & 0 & 0   \\[0.8mm]\hline \\[-3mm]
0  &   &    &   &        \\[0.8mm]
0  & \multicolumn{4}{c}{\raisebox{-10pt}[0pt][0pt]{\Large $U_L$}}\\[0.8mm]
0  &      &          & &             \\[0.8mm]
0  &      &          & &
\end{array}
\right],
\end{align}
where $4 \times 4$ submatrix $U_L$ is consisting of the coefficients of 
the linear combinations mentioned above.
From this reduced matrix, we have energies of the $\Gamma$- and L-states as
\begin{subequations}
\begin{align}
\varepsilon_\Gamma 
&=\frac{\left(\tilde{\varepsilon}^{0}_\Gamma+\tilde{\varepsilon}^{0}_L- 3|\tilde{V}_{L L}|\right)
-\sqrt{\left(\tilde{\varepsilon}^{0}_\Gamma-\tilde{\varepsilon}^{0}_L+3|\tilde{V}_{L L}|\right)^2
+|\tilde{V}_{\Gamma L} |^2}}{2},\\
%
\varepsilon_L^-
&=\frac{\left(\tilde{\varepsilon}^{0}_\Gamma+\tilde{\varepsilon}^{0}_L-3|\tilde{V}_{L L}|\right)
+\sqrt{\left(\tilde{\varepsilon}^{0}_\Gamma-\tilde{\varepsilon}^{0}_L+3|\tilde{V}_{L L}|\right)^2
+|\tilde{V}_{\Gamma L} |^2}}{2},\\
%
\varepsilon_L^+ 
&=\tilde{\varepsilon}^{0}_L + |\tilde{V}_{L L}|. 
\end{align}
\end{subequations}
As seen in eqs.~(\ref{eq:A8}) and (\ref{eq:A9}),
 the elements of the reduced matrix $\tilde{\varepsilon}^{0}_{\Gamma}$ etc. 
depend on $E$.
We evaluated the elements as follows:
In evaluating band egde energy $\varepsilon_\Gamma$, we set $E=\varepsilon^{0}_\Gamma$ and
in evaluating L-point energy $\varepsilon_L^{\pm}$, we set $E=\varepsilon^{0}_L$.

\begin{figure}
\includegraphics[width=8cm]{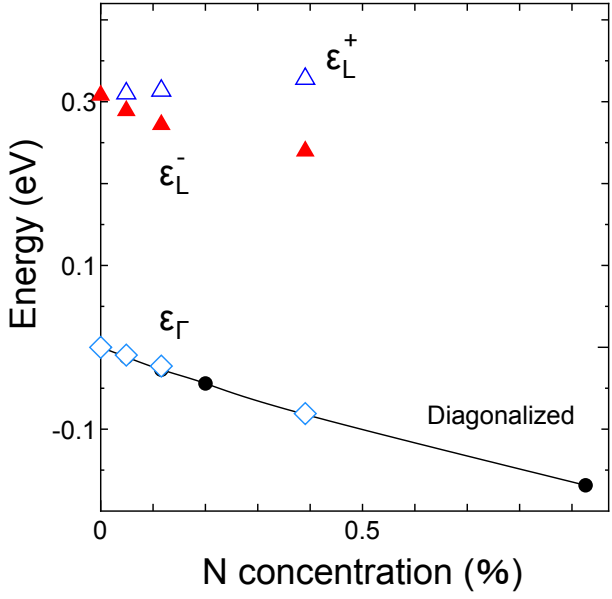}
\caption{(Color Online) Energies calculated by the reduced $\Gamma$-L Hamiltonian.
$\varepsilon_\Gamma$, $\varepsilon_L^-$ and $\varepsilon_L^+$ evaluated by eqs.~(16a) -- (16c) 
are plotted by square, filled triangle, and open triangle, respectively.
Filled circles show energies evaluated by diagonalization of full Hamiltonian, which was 
already shown in Fig.~\ref{fig:6}. }
\label{fig:7}
\end{figure}

In Fig.~\ref{fig:7}, we show energies calculated  by the reduced Hamiltonian.
$\varepsilon_\Gamma$ and $\varepsilon_L^{\pm}$ are plotted by squares and triangles, respectively.
For comparison, we also plot the energies evaluated by diagonalization which were 
already shown in Fig.~\ref{fig:6}.
For the $\Gamma$-state, the two methods give rise to almost the same results.
This good agreement is indicate validity of renormalization procedure.

As for the L-state energy, behavior of $\varepsilon_L^-$ is similar to that of 
 $E_0+\Delta_0$ transition \cite{Tan}.
$\varepsilon_L^+$ seems corresponding to the $E_+$ transition.\cite{Tan,Timoshevskii,Francoeur}
See, for example, Fig.~4 of Ref.~[6].
However, further verification is necessary to apply the present theory to 
high energy states of GaNAs.
In the present paper, we derived $\Gamma$L-reduced Hamiltonian, paying attention mainly to 
band gap reduction.
However, many levels are observed in high energies in GaNAs,\cite{Francoeur} 
and thus only $\Gamma$ and L might be insufficient for description of high energy states. 
It is possible to construct $\Gamma$X- or $\Gamma$XL-reduced Hamiltonian in the same way.
To investigate higher states, inclusion of the X-state would be necessary.

Band calculations using supercell can treat high energy states.
For example, in Ref.~[21] where the first principle calculations were carried out,
the $E_{+}$ transition is assigned to transition to the L-state.\cite{Timoshevskii}
In such calculations, however, we have a difficulty in picking up the state under attention, 
because of a number of states accumulated in energy due to multiply folded bands.
In addition, as we have noted, some L(X)-states mix with the $\Gamma$-state and some do not,
resulting in different dependence on N concentration.
It is not straight forward to investigate high energy states by the supercell calculations.
On the other hand, in the present theory, we can easily obtain high energy states.

\begin{figure}
\includegraphics[width=8cm]{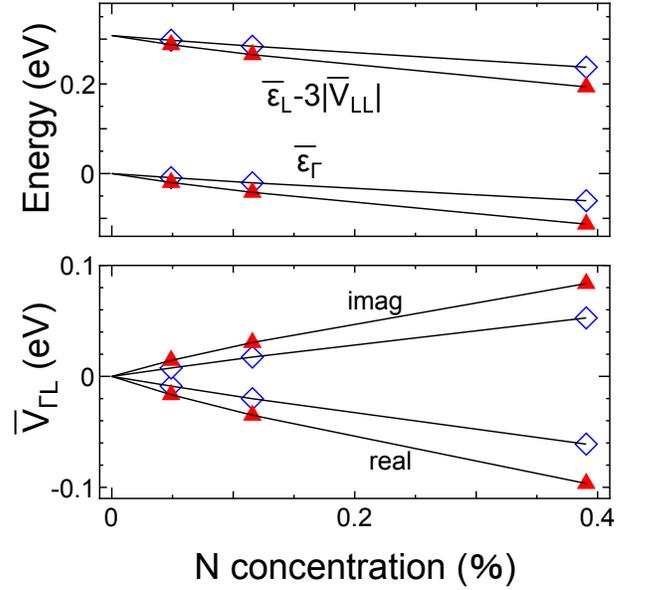}
\caption{(Color Online)  Elements of reduced Hamiltonian are plotted as functions of nitrogen concentration. 
In the upper panel, $\varepsilon^{0}_\Gamma$ and $\varepsilon^{0}_L$ are plotted.
In the lower panel, real and imaginary parts of $V_{\Gamma L}$ are plotted.
Squares (triangles) are values to evaluate $\varepsilon^{0}_\Gamma$ ($\varepsilon^{0}_\Gamma$). 
Note that when nitrogen concentration is zero, these quantities take the values of bulk GaAs.}
\label{fig:8}
\end{figure}

In Fig.~\ref{fig:8}, we show elements of the reduced Hamiltonian.
$\tilde{\varepsilon}^{0}_\Gamma$ and $\tilde{\varepsilon}^{0}_L -3|\tilde{V}_{\Gamma L}|$ are plotted 			in the upper panel and 
 $\tilde{V}_{\Gamma L}$ is plotted in the lower panel as functions of nitrogen concentration.
The squares show the values for $E=\varepsilon^{0}_\Gamma$ used 
to evaluate $\Gamma$-state energy,
whereas triangles show the value with for $E=\varepsilon^{0}_L$ for L-state calculation.
Present scheme where $\Gamma$- and L-states are retained as bases of the reduced Hamiltonian is 
inapplicable when $N$ the supercell dimension is an odd number, because 
the L-point is not included in the set of $\bm k$-points necessary 
for perturbation calculation (see Fig.~\ref{fig:5}).
Thus, calculations were carried out only for $N=4, 6$ and 8 as shown by the squares and triangles 
in Fig.~\ref{fig:7}.
Fortunately, as shown in Fig.~\ref{fig:8}, 
dependence of the matrix elements on nitrogen concentration is nearly linear, 
we can interpolate the matrix elements for calculating wide range of nitrogen concentrations.

\section{Conclusions}

We presented a model to describe the conduction band of dilute nitride compound GaNAs.
Using wavefunctions of conduction band of bulk GaAs as bases, 
we carried out perturbation calculations.
Calculated perturbation matrix elements show that $\Gamma$-L mixing and/or
 $\Gamma$-X mixing are impoprtant for the  band gap reduction.
Though conventional second order formula yields a poor result, diagonalization of the full
Hamiltonian matrix reveals that the present method brings about reasonable results.
By remaining $\Gamma$- and L-states, we renormalized other states to derive effective $2\times 2$ 
reduced Hamiltonian, which also describes well band gap reduction due to nitrogen.

\appendix

\section{Renomalization procedure to reduce Hamiltonian}

We show the procedure to reduce size of the  Hamiltonian by renormalizing 
states whose interaction with the $\Gamma$-state is weak. \cite{Loedin}
First, we divide  basis functions into two groups: 
(A) states to be bases of the reduced Hamiltonian, and (B) the others.
The states of the group (A) are those that interact with $\Gamma$  strongly.
By rearranging order of the bases, 
we rewrite the  matrix ${\cal H}_{\bm k,\bm k'}^{(N)}$ in the form
\begin{align}
{\cal H}_{\bm k,\bm k'}^{(N)}
=
\begin{bmatrix}
{\cal H}_{A}   & {\cal H}_{AB}  \\[2mm]
{\cal H}_{BA}  & {\cal H}_{B}
\end{bmatrix},
\label{eq:A1}
\end{align}
where ${\cal H}_{A}$ is a matrix consisting of the states belonging to the group (A), and so on. 
We omitted the indices $\bm k$, $\bm k'$ and $(N)$ in the right hand side for simplicity.

The secular equation 
is then written as 
\begin{align}
\begin{bmatrix}
{\cal H}_{A} -E{\bf 1}  & \!\!{\cal H}_{AB}  \\[2mm]
{\cal H}_{BA}  & {\cal H}_{B}-E{\bf 1}
\end{bmatrix}
\begin{bmatrix}
c_{A}  \\[2mm]
c_{B}
\end{bmatrix}
=
\begin{bmatrix}
{0}  \\[2mm]
{0}
\end{bmatrix},
\label{eq:A2}
\end{align}
where {\bf 1} is a unit matrix,  {0} a zero vector,
and $c_A$ and $c_B$ are column vectors with corresponding size.
Owing to the choice of states for the groups (A) and (B), 
we expect that elements of 
the matrices ${\cal H}_{B}$, ${\cal H}_{AB}$ and ${\cal H}_{BA}$ are small,
so that we can treat these quantities within lower order terms of expansion series.

Multiplying block by block, eq.~(\ref{eq:A2}) is written as
\begin{align}
({\cal H}_{A} -E{\bf 1})\,c_A  + {\cal H}_{AB}\ c_B={0},  
\label{eq:A3}
\end{align}
and
\begin{align}
{\cal H}_{BA}\ c_A + ({\cal H}^{D}_{B}-E{\bf 1})\, c_B +{\cal H}^{O}_{B}\ c_B ={0},
\label{eq:A4}
\end{align}
where ${\cal H}^{D}_{B}$ and ${\cal H}^{O}_{B}$ denote diagonal and off-diagonal 
parts of ${\cal H}_{B}$, respectively.
We rewrite eq.~(\ref{eq:A4}) in the form 
\begin{align}
c_B = (E{\bf 1}-{\cal H}^{D}_{B})^{-1}{\cal H}_{BA}\ c_A 
+ (E{\bf 1}-{\cal H}^{D}_{B})^{-1}{\cal H}^{O}_{B}\ c_B. 
\label{eq:A5}
\end{align}
This expression enables us to calculate $c_B$ recursively.
The lowest order expression for $c_B$ is readily obtained by setting $c_B = 0$ in the right hand side.
Setting $c_B =0$ after replacing $c_B$ in the right hand by the right hand side itself, 
we have the second expression to $c_B$ as 
\begin{align}
c_B &= (E{\bf 1}-{\cal H}^{D}_{B})^{-1}{\cal H}_{BA}\,c_A \notag \\
    &+ (E{\bf 1}-{\cal H}^{D}_{B})^{-1}{\cal H}^{O}_{B}
        (E{\bf 1}-{\cal H}^{D}_{B})^{-1}{\cal H}_{BA}\,c_A.
\label{eq:A6}
\end{align}
In this way, we can express $c_B$ in a series  until suitable  accuracy is obtained.
Once we have $c_B$, by inserting the expression of $c_B$ into eq.~(\ref{eq:A3}),
we have a secular equation 
\begin{align}
&\left[ {\cal H}_{A} 
+ {\cal H}_{AB}(E{\bf 1}-{\cal H}^{D}_{B})^{-1}{\cal H}_{BA} \right.\notag \\
&\left.       +{\cal H}_{AB}(E{\bf 1}-{\cal H}^{D}_{B})^{-1}{\cal H}^{O}_{B}
(E{\bf 1}-{\cal H}^{D}_{B})^{-1}{\cal H}_{BA} \cdots -E{\bf 1}\right]c_A ={0}.  
\label{eq:A7}
\end{align}
Dimension of the matrices in this equation is that of the group (A).
We thus have the reduced Hamiltonian  
\begin{align}
\tilde{\cal H}_{\bm k,\bm k'}^{(N)}   = {\cal H}_{A} &+ {\cal H}_{AB}
               \left[g(E)+ g(E)\,{\cal H}^{O}_{B}\,g(E)  \right. \notag \\
& \left. + g(E)\,{\cal H}^{O}_{B}\,g(E)\,{\cal H}^{O}_{B}\,g(E) \cdots \right]{\cal H}_{BA},
\label{eq:A8}
\end{align}
with
\begin{align}
g(E)=(E{\bf 1}-{\cal H}^{D}_{B})^{-1},
\label{eq:A9}
\end{align}
in which effects from states of group (B) are effectively contained.

\end{document}